\documentclass[conference]{IEEEtran}
\IEEEoverridecommandlockouts
\usepackage{cite}
\usepackage{amsmath,amssymb,amsfonts}
\usepackage{algorithmic}

\usepackage{caption}
\usepackage{subcaption}
\usepackage{xcolor}
\usepackage{bm}

\usepackage{hyperref}
\usepackage[ruled]{algorithm2e}
\usepackage{multirow}
\usepackage{caption}
\usepackage{tabularx,booktabs}
\usepackage{subcaption}
\usepackage{graphicx}
\usepackage{caption}
\def\BibTeX{{\rm B\kern-.05em{\sc i\kern-.025em b}\kern-.08em
    T\kern-.1667em\lower.7ex\hbox{E}\kern-.125emX}}
\begin{document}

\title{On the Incorporation of Stability Constraints into Sequential Operational Scheduling

}

\author{\IEEEauthorblockN{Wangkun Xu}
\IEEEauthorblockA{
\textit{Imperial College London}\\
London, UK \\
wangkun.xu18@imperial.ac.uk}
\and
\IEEEauthorblockN{Zhongda Chu}
\IEEEauthorblockA{
\textit{Imperial College London}\\
London, UK \\
z.chu18@imperial.ac.uk}
\and
\IEEEauthorblockN{ Florin Capitanescu}
\IEEEauthorblockA{
\textit{Luxembourg Institute of Science and }\\
\textit{Technology}, Luxembourg \\
florin.capitanescu@list.lu}
\and
\IEEEauthorblockN{Fei Teng}
\IEEEauthorblockA{
\textit{Imperial College London}\\
London, UK \\
f.teng@imperial.ac.uk}
}

\maketitle

\begin{abstract}
With the increasing penetration of Inverter-Based Resources (IBRs), power system stability constraints must be incorporated into the operational framework, transforming it into stability-constrained optimization. Currently, there exist parallel research efforts on developing the stability constraints within DC power flow-based unit commitment (UC) and AC Optimal Power Flow (OPF). However, few studies discuss how including such constraints can interact with each other and eventually impact grid stability. In this context, this work simulates a realistic power system decision making framework and provides a thorough analysis on the necessity of incorporating frequency nadir and small signal stability constraints into these sequentially connected two operation stages. The simulation results demonstrate that including both stability constraints in the UC is essential to maintain power system stability, while the inclusion in AC OPF can further improve the stability index.


\end{abstract}

\begin{IEEEkeywords}
Frequency nadir constraints, inverter-based resources, optimal power flow, small signal stability, unit commitment.
\end{IEEEkeywords}

\section{Introduction}

Power system operations consist of a sequence of tasks that are aimed at improving system security and economic efficiency. These tasks vary across time scales: for example, unit commitment, typically implemented with DC power flow, is conducted on a day-ahead basis to determine an optimal status of synchronous generators (SGs) and establish adequate reserves, while AC OPF is solved at minute-level intervals to satisfy constraints in real time \cite{conejo2018power}. To achieve the carbon-neutral target, renewable generation has been massively integrated into the power grid in the last decades. However, the intermittent nature of renewable sources and the unique characteristics of their interfaces—often power electronic devices—pose significant challenges for system operation, security, and stability \cite{8450880}.

To address these issues, substantial efforts have been dedicated to developing stability constraints and embedding them within system scheduling models \cite{bellizio2023transition}. Nonetheless, most existing studies focus on integrating stability constraints into an exact single problem of either UC or OPF, depending on the primary factors that can influence specific stability issues \cite{zhang2023data}.

For frequency issues, most of the existing work encodes the stability constraint into UC as the frequency indices are mostly determined by the system inertia. The authors of \cite{zhang2022frequency} present a frequency security-constrained UC that enables the provision of frequency support and reserve from wind farms. This method accurately quantifies the frequency support capability of wind farms considering the actual grid-connected wind turbine capacity and the wake effect. The concept of frequency security margin is proposed in \cite{zhang2020modeling}, which is incorporated by piecewise linearization. In \cite{paturet2020stochastic}, a frequency-constrained stochastic UC model is presented, where the frequency metrics are derived analytically and then linearized using a boundary extraction approach to maintain the mixed-integer linear formulation. In an integrated electricity-gas system, \cite{yang2022distributionally} includes both frequency stability constraints and natural gas system operational constraints, addressing wind power uncertainty by distributional robustness.

The small-signal (SS) stability issues of highly IBR-penetrated systems are mostly discussed in AC OPF. For example, \cite{liu2021explicit} proposes a data-driven small-signal stability constrained optimal power flow method with high computational efficiency, where the support vector machine with a kernel function is used to derive the explicit data-driven surrogate constraint for small-signal stability. The authors in \cite{pullaguram2021small} formulate a similar problem in inverter-based AC micro-grids. To tackle the non-convexity due to the presence of the nonlinear stability constraint, two distinct convex relaxation approaches, namely semidefinite programming and parabolic relaxations are further developed. Furthermore, novel semidefinite programming has been developed in \cite{pareek2021convexification} where the objective penalization is utilized for feasibility recovery, making the method computationally efficient for large-scale systems.

Despite the distinct literature on the encoding stability constraints in UC and AC OPF, it is unclear whether such constraints need to be included in both stages or only one of them. For the stability-constrained UC problem, the network is mostly modeled by the linearized DC power flow or even neglected. As a consequence, the optimal solution that admits the stability constraints may not ensure the system stability in the actual system where the network is characterized by the nonlinear AC power flow. 

In contrast, for the stability-constrained AC OPF problem, it is typically assumed that all the SGs in the system are online and the main variables that influence the system stability are the active and reactive power set points of the generators or the curtailment of the renewable. However, if the status of the SGs in the system is determined in a conventional fashion (without any stability constraints), solely depending on the generation (re-)dispatch may not be able to maintain the system stability cost-efficiently or even cause infeasibility issues under certain circumstances where more inertia or system strength should have been provided from SGs. 

\begin{figure}[!t]
    \centering
    \includegraphics[width=0.95\linewidth]{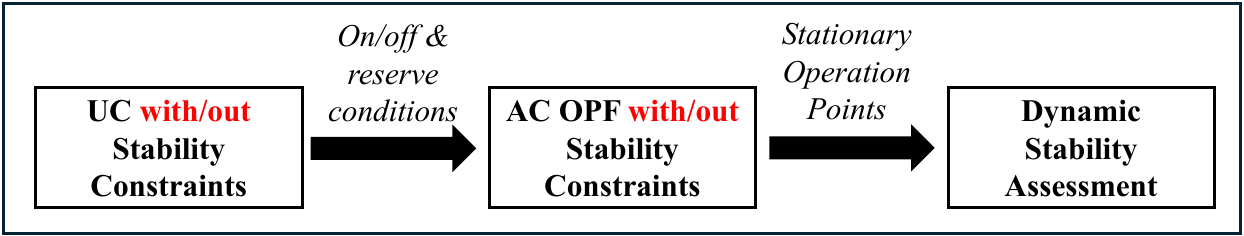}
    \caption{Sequential power system operations with and without stability considerations. This paper demonstrates the necessity of including stability constraints at both stages.}
    \label{fig:flowchart}
\end{figure}

In this context, this paper aims to provide a thorough analysis of the need to incorporate different types of stability constraints such as frequency nadir and small signal stability into sequential UC and AC OPF. Referring to realistic power system operation settings, the stability and economic performances will be evaluated on the operating point determined by AC OPF. See Fig. \ref{fig:flowchart} for an illustration. 


\section{Problem Set-up}

In this paper, the bold letters represent the vectors. $\mathcal{N}_g$, $\mathcal{N}_r$, and $\mathcal{N}_l$ are the set of SGs, IBRs, and loads. All IBRs are assumed to operate in grid following (GFL) mode. The UC and AC OPF are formulated in a way similar to \cite{conejo2018power}. Based on renewable and load profile $\bm{p}_r$ and $\bm{p}_l$, the day-ahead UC can be compactly written as a function $\{\bm{u}_g, \bm{p}_g^{\text{uc}}, \bm{\theta}^{\text{uc}}\}=\mathcal{UC}(\bm{p}_l, \bm{p}_r)$ where $\bm{u}_g$, $\bm{p}_g^{\text{uc}}$ and $\bm{\theta}^{\text{uc}}$ are the vectors of the generator on/off status, generator set point, and nodal voltage angle. Taking the generator scheduling as input, the AC OPF is denoted as $\{\bm{p}_g^{\text{opf}}, \bm{p}_r^{\text{opf}}, \bm{q}_g^{\text{opf}}, \bm{v}^{\text{opf}}, \bm{\theta}^{\text{opf}}\} = \mathcal{OPF}(\bm{u}_g, \bm{p}_g^{\text{uc}}, \bm{p}_l, \bm{p}_r)$ where the decision variables include active and reactive power of the generators, active power of renewable after curtailment, as well as voltage magnitude and angle. To demonstrate the necessity of including stability constraints into \textbf{both stages}, this paper takes into account regular $N-1$ active power reserve, frequency stability, and small-signal stability constraints \cite{chu2023stability}. 


\subsection{Generator Active Power Reserve Constraints}

In addition to the regular UC and AC OPF formulation \cite{conejo2018power}, we include active power reserve constraint to support the $N-1$ generator shortage contingency, i.e.,
\begin{equation} \label{eq:reserve}
    \sum_{i\in\mathcal{N}_g\setminus\{j\}} \bm{u}_{g,i}\bm{p}_{g,i}^\text{max} - \sum_{i\in\mathcal{N}_l}\bm{p}_{l,i} + \sum_{i\in\mathcal{N}_{r}}\bm{p}_{r,i} \geq 0, \; \forall j\in\mathcal{N}_g
\end{equation}
Note that when $j$ is an offline generator, \eqref{eq:reserve} is trivially satisfied with equality.



\subsection{Frequency Stability Constraints}

\subsubsection{Frequency Nadir Index}

Due to the decrease of system inertia, it is essential to consider the frequency stability constraints in the operation \cite{xu2024efficient}. Under the center of inertia (CoI) setting, frequency dynamics in a multi-machine power system can be expressed in the form of a single swing equation:
\begin{equation*}
    \label{sw1}
    2H\frac{d\Delta f(t)}{d t} = -D \Delta f(t) + \Delta R(t) -\Delta P,
\end{equation*}
where $H$ is the CoI constant of the system;  $D$ is the system damping; $\Delta P$ is a step disturbance at $t=0$. In the case where the primary frequency response is provided by the SGs, $\Delta R(t)$ can be represented according to the following scheme \cite{6714513}:
\begin{equation*}
\label{R}
\Delta R(t)=
     \begin{cases}
       \frac{R}{T_d}t, & \; 0\le t< T_d \\ 
       R, & \; T_d\le t
     \end{cases},
\end{equation*}
where $R$ is the reserve of the synchronous generators from the primary frequency response (PFR); $T_d$ is the reserve delivery time. Accordingly, the frequency nadir constraint can be analytically derived as in \cite{chu2020frequency}:
\begin{equation} \label{eq:frequency}
    H\cdot R \geq \frac{\Delta P^2 T_d}{4\Delta f_\text{lim}},
\end{equation}
where $\Delta f_\text{lim}$ is the maximum nadir requirement. To simplify the analysis, the system damping is ignored and \eqref{eq:frequency} becomes conservative. 

\subsubsection{Constraint Formulation}

Similarly to \eqref{eq:reserve}, we consider $N-1$ generator contingency where $\Delta P$ is considered as the output of any of the online generators. Therefore \eqref{eq:frequency} becomes
\begin{subequations}\label{eq:fsc}
    \begin{equation}\label{eq:fsc_1}
        \mathcal{H}_j \mathcal{R}_j \geq \frac{\bm{p}_{g,j}^2 T_d}{4\Delta f_\text{lim}} \quad \forall j\in\mathcal{N}_g, \text{ with}
    \end{equation}
    \begin{equation}\label{eq:fsc_2}
        \mathcal{H}_j  = \sum_{i\in\mathcal{N}_g\setminus \{j\}} \bm{u}_{g,i}\bm{H}_i
    \end{equation}
    \begin{equation}\label{eq:fsc_3}
        \mathcal{R}_j  = \sum_{i\in\mathcal{N}_g\setminus \{j\}} \bm{R}_i
    \end{equation}
\end{subequations}
where $\bm{H}$ is a vector of SG inertia constants and $\bm{R}$ is the PFR of each generator as decision variable.

Note that when $\bm{u}_{g,j}=0$, e.g., an offline generator is removed, $\bm{p}_{g,j}=0$ and \eqref{eq:fsc} is trivially satisfied. Meanwhile the PFR should also be limited as
\begin{equation}\label{eq:pfr}
    0 \leq \bm{R}_i \leq \min\left\{\bm{u}_{g,i}\bm{p}_{g,i}^\text{max} - \bm{p}_{g,i}, \bm{R}_i^{\text{max}} \right\}, \quad i \in \mathcal{N}_g
\end{equation}
where $\bm{R}^\text{max}$ is the maximum active power response that can be delivered, determined by the physical characteristic of SGs.

In addition, for $\forall j\in\mathcal{N}_g$, \eqref{eq:fsc_1} can be equivalently written in the form of disciplined convex programming,
\begin{equation} \label{eq:fsc_soc}
    \mathcal{H}_j + \mathcal{R}_j \geq \left\|\begin{bmatrix}
        \sqrt{\frac{T_d}{\Delta f_\text{lim}}} \bm{p}_{g,j} \\
        \bm{H}_j - \bm{R}_j
    \end{bmatrix}\right\|_2
\end{equation}
which is a second-order cone constraint on $\bm{u}_{g,i}$ and $\bm{p}_{g,i}$.

Consequently, the nadir constraint, represented by \eqref{eq:fsc_soc}, \eqref{eq:fsc_2}, \eqref{eq:fsc_3} and \eqref{eq:pfr}, can be included in both UC and AC OPF.

\subsection{Small Signal Stability}
\subsubsection{Stability Index}
In weak grids (large grid impedance), the IBR PCC voltage can be significantly affected by the current injection into the grid, forming a self-synchronization loop (positive feedback) and hence undermining the GFL IBR synchronization stability. The method proposed in \cite{8488538,gOSCR} is utilized here where the small signal stability of PLL-based GFL IBRs is assessed through the generalized short circuit ratio (gSCR). The small-signal synchronization stability is dominated by the dynamics of IBRs and the network (dominated by the admittance matrix) with which they are connected. Therefore, the closed-loop linearized system dynamic is first built upon the IBR and network dynamics. Consequently, the small-signal stability analysis is carried out based on the linear control theory. 

In detail, the definition of $Y_{eq}$ is given by:
\begin{equation}\label{eq:small_signal_constraint_1}
\begin{aligned}
    \mathrm{gSCR} &= \lambda_{\mathrm{min}} (\bm{Y}_{eq}) \\
    \bm{Y}_{eq} &= \mathrm{diag}\left(\frac{\bm{v}^2_{r}}{\bm{p}_{r}}\right) \bm{Y}_{red},
\end{aligned}
\end{equation}
where $\mathrm{diag}\left({\bm{v}^2_{r}}/{\bm{p}_{r}}\right)$ is the diagonal matrix of the GFL IBR terminal voltage $\bm{v}_{r}$ and output power $\bm{p}_{r}$; $\bm{v}_r^2 / \bm{p}_r$ should be understand as element-wise division; $\bm{Y}_{red}$ is the reduced node admittance matrix after eliminating passive buses and infinite buses. Note that $\bm{Y}_{eq}$ is diagonalizable with its smallest eigenvalue $\lambda_{\mathrm{min}} (\bm{Y}_{eq})\in \mathbb{R}^+$ representing the connectivity of the network, and thus the grid voltage strength and the small-signal synchronization stability constraint can be formulated as \cite{gOSCR}:
\begin{equation}
\label{eq:gSCR_cstrt}
    \mathrm{gSCR} \ge \mathrm{gSCR}_{\mathrm{lim}},
\end{equation}
where $\mathrm{gSCR}_{\mathrm{lim}}$ is the critical (minimum) gSCR that needs to be maintained to ensure the small signal stability of the GFL units. Furthermore, based on the assumption that the system voltages stay close to $1\,\mathrm{p.u.}$ during normal operation and small disturbances, the critical gSCR is an operation-independent value, which can be determined offline with or without the detailed control parameters of the gird-following IBRs \cite{gOSCR}.

The above small-signal setting is verified as follows. First, it accounts for the impact of GFL and the networks. The GFL dynamics influences the value of $\mathrm{gSCR}_{\mathrm{lim}}$, whereas the GFL capacity and location as well as the network influence the value of $\mathrm{gSCR}$, by influencing $\bm{Y}_{eq}$. In other words, the gSCR-based stability analysis decouples the dynamics-related quantities with the steady-state quantities ($\bm{Y}_{eq}$). The former influences $\mathrm{gSCR}_{\mathrm{lim}}$ and is determined by the GFL and network dynamics, which are fixed at the system operation stage once the GFL control algorithm and parameters are selected. The latter influences $\mathrm{gSCR}$ and is determined by the GFL capacity and location as well as the admittance matrix. As a result, by forcing the $\mathrm{gSCR}\le \mathrm{gSCR}_{\mathrm{lim}}$ during system operation, the small signal stability can be maintained.

Second, although the derivation in \cite{8488538,gOSCR} does not include SGs explicitly, it includes an arbitrary number of infinite buses (ideal voltage sources), which increase the grid strength in the system. Since an SG can be modeled as a voltage source behind an impedance for small signal analysis due to its voltage source behaviors, the impact of SGs can be modeled by augmenting the system admittance matrix with the internal impedance of SGs \cite{xin2022many}. By doing so, a general multi-machine system can be converted to the formulation in \cite{8488538,gOSCR}.

\subsubsection{Constraint Formulation}
As demonstrated above, the critical $\mathrm{gSCR}_{\mathrm{lim}}$ in \eqref{eq:gSCR_cstrt} is fixed during the system scheduling stage, whereas the gSCR is dependent on the generator status and renewable energy output, i.e., $\mathrm{gSCR} = \mathrm{gSCR} (\bm{p}_r,\bm{u}_{g})$. This dependence is explicitly derived in this section. 


To start, let $\bm{Y}_0$ represent the nodal admittance matrix with pure reactive transmission line \cite{gOSCR}, an augmented admittance matrix with the reactance of SGs included is considered,
\begin{equation*}
\label{eq:Y1}
    \bm{Y} = \bm{Y}_0 + \bm{Y}_g,
\end{equation*}
During normal operation and small disturbances, SGs can be viewed as voltage sources behind impedances. Therefore, depending on the operating conditions of the SGs, the elements in $\bm{Y}_g$ can be expressed as:
\begin{equation*}
\label{eq:Y2}
    \bm{Y}_{g, ij}=
    \begin{cases}
    \frac{1}{\bm{x}_{g,i}}\bm{u}_{g,i}\;\;&\mathrm{if}\,i = j \in \mathcal{N}_g \\
    0\;\;& \mathrm{otherwise},
    \end{cases}
\end{equation*}
where $\bm{x_}{g}$ is the SG reactance. After permuting $\bm{Y}$ into following block matrix,
\begin{equation*}
\label{eq:Y0}
    \bm{Y}=\begin{bmatrix}
        \begin{array}{c|c}
        \bm{Y}_{\mathcal{N}_r\mathcal{N}_r} & \bm{Y}_{\mathcal{N}_r \delta} \\ \hline
        \bm{Y}_{\delta \mathcal{N}_r} & \bm{Y}_{\delta \delta} 
        \end{array}
    \end{bmatrix},
\end{equation*}
with $\mathcal{N}_r\subset\mathcal{N}$ being the set of GFL IBR nodes and $\delta = \mathcal{N}\setminus \mathcal{N}_r$ being the set of the remaining nodes, $\bm{Y}_{red}$ in \eqref{eq:small_signal_constraint_1} can be finally expressed by the kron reduction: 
\begin{equation}
\label{Y_red}
    \bm{Y}_{red} = \bm{Y}_{\mathcal{N}_r\mathcal{N}_r}- \bm{Y}_{\mathcal{N}_r \delta} \bm{Y}_{\delta \delta}^{-1} \bm{Y}_{\delta \mathcal{N}_r},
\end{equation}

Plugging \eqref{Y_red} into \eqref{eq:small_signal_constraint_1} gives the relationship between the gSCR and the decision variables $(\bm{p}_{r}, \bm{u}_{g})$ in the system scheduling model where \eqref{eq:gSCR_cstrt} should be satisfied.

\subsubsection{Linear Constraint Generation}

Because the eigenvalue problem is nonconvex, a logistic regression model is trained to classify the stable and unstable samples in terms of small signal stability. Let $\mathcal{D} = \mathcal{D}_{\text{stable}}\cup \mathcal{D}_{\text{unstable}}$ be the index set of the historic data $\{(\bm{x}_k,y_k)\}_{k}$, the logistic regression becomes,
\begin{equation}\label{eq:logistic}
    \max_{\bm{w},b} \; \frac{1}{|\mathcal{D}|}\sum_{k\in\mathcal{D}} \bm{\alpha}_k [\bm{y}_k(\bm{w}^T\bm{x}_k+b) - \log(1+e^{\bm{w}^T\bm{x}_k + b})]
\end{equation}
where $\bm{y}_k\in\{0,1\}$ represents the stable and unstable labels; $\alpha_k$ is used to control the conservative level on balancing the stable and unstable samples. Let $\alpha_\text{unstable}$ be the hyperparameter associated with the unstable samples. When $\alpha_\text{unstable}\rightarrow\infty$, the trained classifier will become conservative in which all unstable data samples must be correctly classified. Equivalently, \eqref{eq:logistic} becomes a convex optimization problem,
\begin{equation*}
    \begin{aligned}
        \min_{\bm{w},b} & \quad \frac{1}{|\mathcal{D_\text{stable}}|} \sum_{k\in\mathcal{D}_\text{stable}} \log (1 + e^{\bm{w}^T\bm{x}_k + b}) \\
        \text{s.t.} & \quad \bm{w}^T\bm{x}_k + b \geq 0, \quad k \in \mathcal{D}_\text{unstable}
    \end{aligned}
\end{equation*}

Given the optimal $\bm{w}^\star,b^\star$, a linear constraint $\bm{w}^{\star,T}\bm{x} + b^\star \leq 0$ can be included in power system optimization as the constraint of small-signal stability. 


\section{Simulations and Discussions}

We use the IEEE 5-machine 14-bus system as a testbed. The static data are taken from the \texttt{PyPower}\footnote{\url{https://github.com/rwl/PYPOWER/blob/master/pypower/case14.py}.} where bus 5, 11, 13, and 14 are equipped with wind farms. The load and wind profiles are modified from the open source Texas Backbone Power System\footnote{\url{https://rpglab.github.io/resources/TX-123BT/}.}. Meanwhile, the dynamic configuration is given in Table \ref{tab:dynamic_conf}. We consider sequential UC and AC OPF decision makings as shown in Fig. \ref{fig:flowchart}. UC is compiled by \texttt{CVXPY} and solved by \texttt{GUROBI} with MIPGap set as 0.01\%. The AC OPF is solved by \texttt{CYIPOPT} with acceptable\_tol equal to 0.01\%. Remaining solver options are set as default. We evaluated the frequency nadir based on the dispatch result after AC OPF using \eqref{eq:fsc} (or equivalently \eqref{eq:fsc_soc}). The stability of the small signal is evaluated by the original definition of gSCR as in \eqref{eq:small_signal_constraint_1}.

\begin{table}[]
    \centering
    \footnotesize
    \caption{Generator Dynamic Configurations.}
    \begin{tabular}{c|c|c}
       & \textbf{Meaning} & \textbf{Parameter} \\\hline
       $\Delta f_\text{lim}$ & nadir limit & 0.8 Hz \\
       $T_d$ & PFR time & 10 s \\
       $\bm{H}$ & Inertia constant & [0.8, 1.0, 3.0, 2.0, 0.6] s \\
       $\bm{x}_d^\prime$ & SG transient reactance & [0.10, 0.13, 0.2, 0.16, 0.12] p.u. \\
       $\bm{R}^\text{max}$ & Maximum PFR & [0.4, 0.3, 0.25, 0.20, 0.35] p.u. \\
       $\text{gSCR}_\text{lim}$ & Minimum gSCR & 2.5
    \end{tabular}
    \label{tab:dynamic_conf}
\end{table}


We consider two renewable penetration levels defined as the Renewable-to-Load (RtL) ratio over a 24-hour operation period. The complete simulation results are summarized in Table \ref{tab:rtl35} and \ref{tab:rtl55} for RtL $35\pm5\%$ and $55\pm5\%$, respectively. For each case, performance is taken into account for 10 random samples. For example, for RtL in $35\pm5\%$ in Table \ref{tab:rtl35}, we randomly sample 10 renewable and load profiles within this range and simulate the UC and AC OPF with different combinations of stability constraints. In detail, \textbf{UC-Nd}, \textbf{OPF-Nd}, \textbf{UC-SS}, and \textbf{OPF-SS} represent if the small-signal (SS) and nadir (Nd) constraints are included in UC and AC OPF. The large penalty for load shedding has been added to the AC OPF and the load-shedding rate \textbf{LSR} records the percentage of occurrence when load shedding is triggered. For the frequency nadir, load shedding occurs when more PFR is needed. For small-signal stability, load shedding is forced to compensate on the curtailed renewable. Therefore, the LSR can be considered as the infeasibility rate of AC OPF. The \textbf{Cost} is the total cost of UC and AC OPF. \textbf{VR} represents the stability violation ratio. Although the stability performances after UC are not the realistic one presented in the system, we include those indices for comparison purposes. 

In general, it can be observed that when neither stability constraints are included, both stability criteria are significantly violated. We will discuss the simulation results from different perspectives. Note that the simulation results can be case-dependent and subject to the formulation of the stability index in this paper. However, the conclusions can be generally meaningful.

\textit{Performance of nadir constraints on nadir index.} Including the nadir constraint in AC OPF alone causes infeasibility (e.g., around 80\% LSR for both RtLs), due to insufficient online SGs committed at the UC stage. In contrast, including the nadir constraint in UC alone can significantly improve but cannot guarantee the nadir after ACOPF, and the constraint violation increases as the RtL increases (e.g., from 35.42\% to 59.17\%). The safest way is to include nadir constraints in both UC and AC OPF. However, load shedding still occurs because UC cannot foresee the different generator dispatch of AC OPF in sequential decision-making. 

\textit{Performance of SS constraints on SS index.} Including the SS constraint in AC OPF alone causes infeasibility. Compared to the frequency nadir, it is more sensitive to the RtL (e.g., the LSR increases from 18.33\% to 48.75\% when RtL increases). Moreover, including the SS constraint in UC alone seems to be sufficient to maintain the SS stability after AC OPF most of the time. This is because if the renewable curtailment is ignored in AC OPF, the gSCR is purely determined by the commitment status of SGs. Furthermore, the 0.83\% and 7.50\% violation of SS stability in AC OPF can be eliminated by renewable curtailment after including SS constraints in AC OPF.

\textit{Performance of nadir constraints on SS index.} When no SS is considered, the SS violation rate for UC and AC OPF are the same, which is purely determined by the nadir constraints in the UC stage. This is because the gSCR is determined solely by the commitment of SGs if there is no renewable curtailment. Meanwhile, the nadir constraints can slightly improve SS stability by 8\% for RtL=$35\pm5\%$ and 15\% for RtL=$55\pm5\%$ as the SGs reactance can contribute to the increase of gSCR.  

\textit{Performance of SS constraints on nadir index.} The inclusion of the SS constraint in UC can contribute to the frequency nadir in both UC and AC OPF by reducing the violation by 25\% when RtL=$35\pm5\%$ and 70\% when RtL=$55\pm5\%$. Note that the nadir improvement is more significant with high RtL. In addition, adding an SS constraint in AC OPF cannot effectively improve the frequency nadir. This is because when the commitment of SGs is fixed (therefore the inertia is fixed), curtailing the renewable cannot effectively influence the PFR. Note that when both stability constraints are included, the LSRs are reduced compared to those with only the nadir constraints included.    

\begin{table*}[h]
    \centering
    \footnotesize
    \caption{Performances of SC-UC and SC-ACOPF Under RtL $35\pm5\%$.}
    \begin{tabular}{|c|c|c|c|c|c|c|c|c|c|c|}
                 \textbf{UC-Nd} & \textbf{OPF-Nd} & \textbf{UC-SS} & \textbf{OPF-SS} & \textbf{LSR} & \textbf{Cost (\$)} & \textbf{NdVR after UC} & \textbf{NdVR after OPF} & \textbf{SSVR after UC} & \textbf{SSVR after OPF}  \\\hline\hline
                 \multicolumn{10}{|c|}{NO STABILITY CONSTRAINTS} \\\hline
                 OFF & OFF & OFF & OFF & 0.00\% & 102.61 & 98.33\% & 98.33\% & 20.00\% & 20.00\% \\\hline
                 \multicolumn{10}{|c|}{WITH FREQUENCY NADIR CONSTRAINTS} \\\hline
                 OFF & ON & OFF & OFF & 80.00\% & 35113.13 & 98.33\% & 0.00\% & 20.00\% & 20.00\%  \\
                 ON & OFF & OFF & OFF & 0.00\% & 112.89 & 0.00\% & 35.42\% & 12.08\% & 12.08\%  \\
                 ON & ON & OFF & OFF & 4.58\% & 274.24 & 0.00\% & 0.00\% & 12.08\% & 12.08\%   \\\hline
                 \multicolumn{10}{|c|}{WITH SMALL SIGNAL STABILITY CONSTRAINTS} \\\hline
                 OFF & OFF & OFF & ON & 18.33\% & 4510.41 & 98.33\% & 99.17\% & 20.00\% & 0.00\%    \\
                 OFF & OFF & ON & OFF & 0.00\% & 110.30 & 73.33\% & 72.50\% & 0.00\% & 0.83\%  \\
                 OFF & OFF & ON & ON &  0.00\% & 117.437  &  73.33\%  &  72.50\%  & 0.00\%  & 0.00\%  \\\hline
                 \multicolumn{10}{|c|}{WITH BOTH STABILITY CONSTRAINTS} \\\hline
                 ON & ON & ON & ON  & 3.75\% & 213.58 &  0.00\%  & 0.00\%  &  0.00\%  &  0.00\%  \\\hline
    \end{tabular}
    \label{tab:rtl35}
\end{table*}

\begin{table*}[h]
    \centering
    \footnotesize
    \caption{Performances of SC-UC and SC-ACOPF Under RtL $55\pm5\%$.}
    \begin{tabular}{|c|c|c|c|c|c|c|c|c|c|c|}
                 \textbf{UC-Nd} & \textbf{OPF-Nd} & \textbf{UC-SS} & \textbf{OPF-SS} & \textbf{LSR} & \textbf{Cost (\$)} & \textbf{NdVR after UC} & \textbf{NdVR after OPF} & \textbf{SSVR after UC} & \textbf{SSVR after OPF}  \\\hline\hline
                 \multicolumn{10}{|c|}{NO STABILITY CONSTRAINTS} \\\hline
                 OFF & OFF & OFF & OFF & 0.00\%  & 84.585  & 91.67\%  &  95.00\%  &  64.17\%  &  64.17\% \\\hline
                 \multicolumn{10}{|c|}{FREQUENCY NADIR CONSTRAINTS} \\\hline
                 OFF & ON & OFF & OFF & 80.83\%  & 34027.85  & 91.67\%  &   0.00\%   &  64.17\%   & 64.17\%  \\
                 ON & OFF & OFF & OFF & 0.00\% & 92.32  &  0.00\%  & 59.17\%  & 49.17\%   &  49.17\%  \\
                 ON & ON & OFF & OFF & 11.25\%  &  578.52  &  0.00\%   &   0.00\%  & 49.17\%  & 49.17\%   \\\hline
                 \multicolumn{10}{|c|}{SMALL SIGNAL STABILITY CONSTRAINTS} \\\hline
                 OFF & OFF & OFF & ON & 48.75\% & 27196.81 & 91.67\%  & 95.42\%  & 64.17\%  &  0.00\%  \\
                 OFF & OFF & ON & OFF & 0.00\%  & 108.17  & 23.33\%   &  23.33\%  & 0.00\%  &    7.50\%  \\
                 OFF & OFF & ON & ON  & 0.00\%  & 153.07  & 23.33\%  & 23.33\%  &  0.00\%   &   0.00\%  \\\hline
                 \multicolumn{10}{|c|}{WITH BOTH STABILITY CONSTRAINTS} \\\hline
                 ON & ON & ON & ON &  1.25\%  & 171.2  & 0.00\%    &   0.00\% &  0.00\%  &  0.00\%   \\\hline
    \end{tabular}
    \label{tab:rtl55}
\end{table*}

\section{Conclusion and Future Work}
This paper addresses a critical gap in the literature by exploring the incorporation of stability constraints into different power system operation stages. To investigate the problem, sequential decision making, including UC and AC OPF, with generator reserve, frequency nadir, and small-signal stability constraints, has been formulated. The main takeaways of this paper include the following: 1) Both nadir and small-signal stability constraints should be included in UC. This will significantly improve the stability indices for the real system. To guarantee stability, further inclusion of the constraint in the AC OPF is necessary; 
2) Including the stability constraint in AC OPF alone can cause infeasibility;
3) Stability indices can benefit each other. As both nadir and small signal stabilities are strongly dependent on the status of SGs, it is viable to seek a conservative formulation where both constraints are included. Future work seeks to investigate more stability indices \cite{chu2023stability} and their inclusion in AC security-constrained OPF \cite{alizadeh2022envisioning}.

\bibliographystyle{IEEEtran}
\bibliography{IEEEabrv,Reference.bib}

\end{document}